\documentclass[showpacs,preprintnumbers,amsmath,amssymb,prx,twocolumn,nofootinbib]{revtex4}

\usepackage{epsfig,verbatim,textcase}

\renewcommand{\vec}[1]{{\bf{#1}}}

\begin{document}

\title{Pioneer Anomaly: Evaluating Newly Recovered Data}

\pacs{44.40.+a,45.20.df,95.10.Ce,95.10.Eg,95.55.-n,95.55.Jz,95.55.Pe,95.75.-z}


\keywords{Pioneer anomaly, deep-space navigation, thermal modeling.}

\author{Viktor T. Toth}
\address{Ottawa, ON  K1N 9H5, Canada}
\homepage{http://www.vttoth.com/}

\author{Slava G. Turyshev}
\affiliation{Jet Propulsion Laboratory, California Institute of Technology\\
4800 Oak Grove Drive, Pasadena, CA 91109-0899, USA}
\email{turyshev@jpl.nasa.gov}

\begin{abstract}
The Pioneer 10/11 spacecraft yielded the most precise navigation in deep space to date. However, their radio-metric tracking data received from the distances between 20--70 astronomical units from the Sun consistently indicated the presence of a small, anomalous, Doppler frequency drift. The drift is a blue frequency shift that can be interpreted as a sunward acceleration of $a_P = (8.74\pm 1.33)\times 10^{-10}$~m/s$^2$ for each particular spacecraft. This signal has become known as the Pioneer anomaly; the nature of this anomaly remains unexplained.

New Pioneer 10 and 11 radio-metric Doppler data recently became available. The much extended set of Pioneer Doppler data is the primary source for new upcoming investigation of the anomaly. We also have almost entire records of flight telemetry files received from the the Pioneers. Together with original project documentation and newly developed software tools, this additional information is now used to reconstruct the engineering history of both spacecraft. To that extent, a thermal model of the Pioneer vehicles is being developed to study possible contribution of thermal recoil force acting on the two spacecraft. In addition, to improve the accuracy of orbital reconstruction, we developed a new approach that uses actual flight telemetry data during trajectory analysis of radio-metric Doppler files. The ultimate goal of these efforts is to investigate possible contributions of the thermal recoil force to the detected anomalous acceleration.

\end{abstract}

\maketitle


\section{Introduction}

The first spacecraft to leave the inner solar system \citep{JPL1998,JPL1999,JPL2002,JPL2005}, Pioneers 10 and 11 were designed to conduct an exploration of the interplanetary medium beyond the orbit of Mars and perform close-up observations of Jupiter during the 1972-73 Jovian opportunities.

The spacecraft were launched in March 1972 (Pioneer 10) and April 1973 (Pioneer 11) on top of identical three-stage Atlas-Centaur launch vehicles. After passing through the asteroid belt, Pioneer 10 reached Jupiter in December 1973. The trajectory of its sister craft, Pioneer 11, in addition to visiting Jupiter in 1974, also included an encounter with Saturn in 1979 (see \citep{JPL2002,MDR2005} for more details).

After the planetary encounters and successful completion of their primary missions, both Pioneers continued to explore the outer solar system. Due to their excellent health and navigational capabilities, the Pioneers were used to search for trans-Neptunian objects and to establish limits on the presence of low-frequency gravitational radiation \citep{PC202}.

Eventually, Pioneer 10 became the first man-made object to leave the solar system, with its official mission ending in March 1997. Since then, NASA's Deep Space Network (DSN) made occasional contact with the spacecraft. The last successful communication from Pioneer 10 was received by the DSN on 27 April 2002. Pioneer 11 sent its last coherent Doppler data in October 1990; the last scientific observations were returned by Pioneer 11 in September 1995.

The orbits of Pioneers 10 and 11 were reconstructed based primarily on radio-metric (Doppler) tracking data. The reconstruction between heliocentric distances of 20--70 AU yielded a persistent small discrepancy between observed and computed values \citep{JPL2002,JPL2005,MDR2005}. After accounting for known systematic effects \citep{JPL2002}, the unmodeled change in the Doppler residual for Pioneer 10 and 11 is equivalent to an approximately sunward constant acceleration of
\[
a_P = (8.74\pm 1.33)\times 10^{-10}~\mathrm{m/s}^2.
\]
The magnitude of this effect remains approximately constant within the 3~dB gain bandwidth of the high-gain antenna (HGA) \cite{JPL2005,MDR2005}. The nature of this anomalous acceleration remains unexplained; this signal has become known as the Pioneer anomaly.

There were numerous attempts in recent years to provide an explanation for the anomalous acceleration of Pioneers 10 and 11. These can be broadly categorized as either invoking conventional mechanisms or utilizing principles of ``new physics''.

Initial efforts to explain the Pioneer anomaly focused on the possibility of on-board systematic forces. While these cannot be conclusively excluded \citep{JPL2002,JPL2005}, the evidence to date does not support these mechanisms: the magnitude of the anomaly exceeds the acceleration that these mechanisms would likely produce, and the temporal evolution of the anomaly differs from that which one would expect, for instance, if the anomaly were due to thermal radiation of a decaying nuclear power source.

Conventional mechanisms external to the spacecraft were also considered. First among these was the possibility that the anomaly may be due to perturbations of the spacecrafts' orbits by as yet unknown objects in the Kuiper belt. Another possibility is that dust in the solar system may exert a drag force, or it may cause a frequency shift, proportional to distance, in the radio signal. These proposals could not produce a model that is consistent with the known properties of the Pioneer anomaly, and may also be in contradiction with the known properties of planetary orbits.

The value of the Pioneer anomaly happens to be approximately $cH_0$, where $c$ is the speed of light and $H_0$ is the Hubble constant at the present epoch. Attempts were made to exploit this numerical coincidence to provide a cosmological explanation for the anomaly, but it has been demonstrated that not only this approach would produce an effect with the opposite sign \citep{JPL2002,MDR2005}, it will also will have a much smaller magnitude.

As the search for a conventional explanation for the anomaly appeared unsuccessful, this provided a motivation to seek an explanation in ``new physics''. No such attempt to date produced a clearly viable mechanism for the anomaly \cite{MDR2005}.

The inability to explain the anomalous behavior of the Pioneers with conventional physics has resulted in a growing discussion about the origin of the detected signal. The limited size of the previously analyzed data set, also limits our current knowledge of the anomaly. We emphasize that in order to determine the origin of $a_P$ and especially before any serious discussion of new physics can take place, one must analyze the entire set of radio-metric Doppler data received from the Pioneers.

In this paper we report on the progress of the recovery of the Pioneers' radio-metric Doppler data and flight telemetry and the status of the analysis of the Pioneer anomaly. The paper is organized as follows. In Section~\ref{sec:recovery} we discuss the recovery of the extended Doppler data set and its current status. In Section~\ref{sec:tel-recovery} we present the flight telemetry and discuss its value for the upcoming investigation. In Section~\ref{sec:rec-force} we discuss modeling the thermal recoil force and present preliminary results of this effort. In Section~\ref{sec:summ} we conclude with a summary and outline the next sets in the study of the Pioneer anomaly.

\section{\label{sec:recovery}Recovery of the extended Doppler data set}

\begin{table}
\caption{Pioneer 10 and 11 radio-metric Doppler data used in previous studies and presently available for new analysis.}
{\small
\centering
\begin{tabular}{|c|c|c|}\hline
\multicolumn{3}{|c|}{Data used in the previous analyses}\\\hline
Spacecraft&Time span& Distances (AU)\\
Pioneer 10 & 03.01.87 -- 22.07.98 & 40.0 -- 70.5\\
Pioneer 11 & 05.01.87 -- 01.10.90 & 22.4 -- 31.7\\\hline\hline
\multicolumn{3}{|c|}{Currently available data}\\\hline
Spacecraft&Time span& Distances (AU)\\
Pioneer 10 & 08.09.73 -- 27.04.02 & 4.56 -- 80.2\\
Pioneer 11 & 10.04.73 -- 11.10.94 & 1.01 -- 41.7\\\hline
\end{tabular}
\label{tb:new-doppler-data}
}
\end{table}

As of October 2007, an effort to recover all archived Pioneer 10 and 11 Doppler data, initiated at JPL in June 2005, has been completed; there is now almost 30 years of Pioneer 10 and 20 years of Pioneer 11 Doppler data, most of which was never used in the investigation of the anomaly (Table~\ref{tb:new-doppler-data}).

The recovery process presented many unanticipated challenges. Uncertainty in spin calibration and missing ramp information are two issues that created a major setback in the data recovery project, as most of the newly recovered data files were not properly conditioned. In the following sections, we address these issues and briefly describe the techniques used for orbital analysis.

\subsection{Spin calibration}

Radio communication with the Pioneer spacecraft was maintained using S-band microwave transmissions. The transmission frequency was $\sim$2.1~GHz. The frequency of the return transmission from the spacecraft was $\sim$2.3~GHz. When the spacecraft's radio communication subsystem was in coherent mode, its receiver and transmitter were coupled, and the ratio of the frequency of the signal transmitted vs. the signal received by the spacecraft was exactly $240/221$. It is this phase-coherent mode of operation that allowed precision Doppler measurements by ground stations, often with mHz accuracy.

At this accuracy, one must take into account the combined effect of the spacecraft's spin and the polarization of the radio signal.

The radio signal transmitted to, and returned by, the Pioneer spacecraft was circularly polarized. The spacecraft was physically oriented such that its high-gain antenna always pointed towards the Earth; the spacecraft was spinning around its spin axis, which approximately coincided with the antenna axis.

As a result, each revolution of the spacecraft added a full cycle to the signal in each direction.

At a nominal spin rate of 4.5 revolutions per minute (rpm), this resulted a shift of 0.15 cycles per second in the return frequency. Knowing the actual spin rate of the spacecraft from telemetry, it is possible to calculate, and account for, the spacecrafts' spin more accurately. When a data file containing radio-metric information has been modified to account for this effect due to spin, it is said to be spin calibrated.

When we assembled historical Pioneer Doppler data files for the first time, alarmingly we found that some of the files were spin-calibrated, whereas other files were not. It soon became evident that, in order to render these files suitable for precision orbit calculations, their spin calibration had to be reconstructed from historical records (see discussion in Sec.~\ref{sec:spin}).

\subsection{Ramp Data}

Several years after the launch of the twin Pioneer spacecraft, the Deep Space Network (DSN) began experimenting with ``ramping'' its transmission frequencies, in order to follow the variable frequency shift that is a result of the changing relative velocities of a distant spacecraft and a station on the Earth's surface, during the course of a data transmission session.

Ramping is characterized by a starting frequency and a ramp rate, in Hz/s. Unfortunately, older data file formats used by the DSN had no provisions for storing this ramp data, therefore this information was stored separately. Further compounding the problem was the fact that ramp data records were often used to store the transmission frequency even when no actual ramping took place (i.e., the ramp rate was 0); the transmission frequency recorded in the Doppler record may or may not have coincided with the actual transmission frequency.

When the recently recovered Doppler records were first assembled and analyzed, a strange ``banding'' appeared in the data, with data points offset by several hundred to several thousands Hz from each other. Missing ramp records were identified as a likely cause for this effect in many cases. In order to reconstruct a usable data set, it was necessary to search historical records that contained the necessary ramping information, or discard records for which no reliable ramp records could be found.

\subsection{Data Conditioning}

We realized that unusual behavior of the recovered raw Doppler data discussed in the previous section was due to the fact that these files were not submitted to a proper radio-metric data conditioning (RMDC) procedure \cite{JPL2002}. This procedure is a standard part of preparing orbit tracking data for a trajectory analysis. As part of the standard process RMDC it includes both spin calibration and application of the correct ramp information to raw Doppler data received from DSN stations.

It turned out that RMDC work was necessary for all Pioneer Archival Tracking Data Files (ATDFs; see discussion in \cite{MDR2005}) that we received from National Space Science Data Center (NSSDC), as all of these files were raw unconditioned ATDFs.

Below is a description of the RMDC process required for making Pioneer ATDFs usable for our investigation.

The process is very detail-oriented, tedious, and labor-intensive. It consists of two phases -- analysis and implementation -- that are discussed below.

\subsubsection{Analysis}

Analysis of tracking data files comprises the following critical steps:

\begin{itemize}
\item Generate a summary of the contents of each ATDF;
\item Separate and print the tracking data, ramp data, and transmitter parameters of each file summary;
\item Compare ramp data to transmitter summaries to identify erroneous/extraneous transmitter on periods, and uplink periods which are not covered by ramp records, then transfer this information to the data summary;
\item Check data summaries for erroneous reference frequencies (receiver and exciter); times when the reference frequencies are being ramped, but there are no corresponding ramp records; erroneous status (validity, transmitter on/off, etc.) flags; erroneous identifier (exciter band, receiver band, Doppler mode, synthesizer, exciter type, receiver type, etc.) flags; erroneous Doppler bias values;
\end{itemize}
Sometimes, the validation process can be an almost point-by-point exercise.

\subsubsection{Implementation}

\begin{figure}
\centering \psfig{file=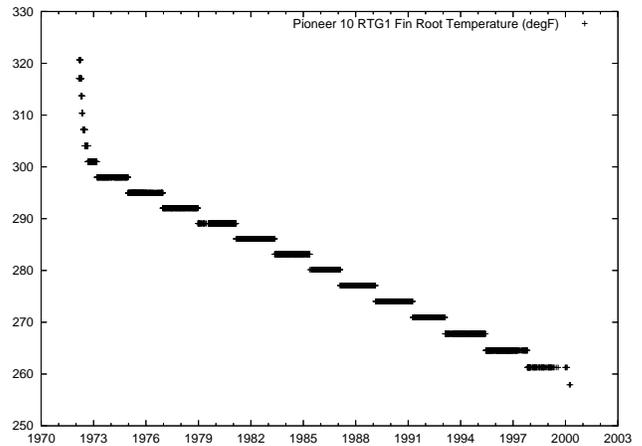, width=\linewidth}
\caption{RTG 1 fin root temperatures ($^\circ$F) for Pioneer 10.}
\label{fig:C201}
\end{figure}

After the analysis of tracking data files has been completed, the necessary processing steps are implemented as follows:

\begin{itemize}
\item Translate the analysis information into change commands (some of the necessary fixes could require as many as 4 commands to implement) to be used with the appropriate program;
\item Execute the program to apply the correction commands (at least 250+ commands for each ATDF);
\item Generate a summary of the updated ATDF;
\item Check the new summaries to confirm that the corrections have been properly applied, and no new problems have been introduced (or ``come to the surface'') by making the changes.
\end{itemize}

Once the steps above are implemented, the tracking data files are ready to be used for generating standard format Orbit Data Files (ODFs) that can be submitted for further analysis with the Orbit Determination Program (ODP) software.

After the two-stage process above is complete, another person takes a ``conditioned'' ODF and runs it through ODP to verify the quality of the file. Experience has shown that this step sometimes identifies further problems, resulting in some additional time and effort being needed before a fully-usable tracking data file is available. If the quality is satisfactory then the file is added to the collection of good files, otherwise the file is returned for further RMDC. The process repeats until all files are found acceptable.

The work on data conditioning of nearly 600+ Pioneer ATDFs is progressing well and will be completed by the end of 2007.
The result of this analysis is that at long last, a usable data set is now available for analysis, covering in particular the early parts of the Pioneer 10 and 11 missions, for which no Doppler data records were available in the past. This is especially important because there has been speculation that there may have been an ``onset'' of the anomalous acceleration of Pioneer 10 and 11 when they reached approximately the orbit of Saturn (see discussion in \cite{MDR2005}). Is this onset a real effect, or just an artifact of incomplete past analyses? If the effect is real, was it caused by, for instance, incorrect solar pressure calibration, or does it represent a hint about the real cause of the Pioneer anomaly? Our upcoming analysis is aimed to answer these and other questions.

\subsection{Orbital analysis with Doppler data}

Doppler data are the recorded measurements of the shift in frequency of the signal received from the spacecraft, as a result of the combined relative motion of the spacecraft and the Earth station(s). The position of the transmitting station changes as a function of time as a result of the combined effects of the Earth's rotation, precession, and orbital motion. The position of the spacecraft can be calculated if its orbital parameters are known. After these calculations, the Doppler shift in frequency can be estimated and compared with the observed frequency shift (for more details on the analysis of radio-metric Doppler data and formats used for these purposes consult \cite{JPL2002,MOYER2000} and references therein).

In principle, an orbit determination algorithm employs a generalized equation of motion (for extensive details, see \cite{MOYER2000}), incorporating not just the Newtonian gravitational effects of massive solar system bodies, but also general relativistic corrections, non-gravitational forces, and equations describing the propagation of the radio signals through the interplanetary medium, the gravitational field of the Sun and planets, and the Earth's atmosphere and ionosphere. If needed, other forces may be incorporated into the model and used for hypothesis testing (see details in \cite{JPL2002}).

The aim of the orbital analysis is to determine the initial position and velocity (state vector) of the spacecraft, as well as the values of additional, model-dependent parameters that yield a best match with observational data. In our case, the goal is to reduce the difference between calculated and measured Doppler values (i.e., the Doppler residual) and then adjust the initial state vector and other parameters until the residual is minimized (for details on models, methods, and formats consult \cite{JPL2002,MOYER2000,Mont-Gill-2005}).

\section{\label{sec:tel-recovery}Flight telemetry}

\begin{figure}
\centering \psfig{file=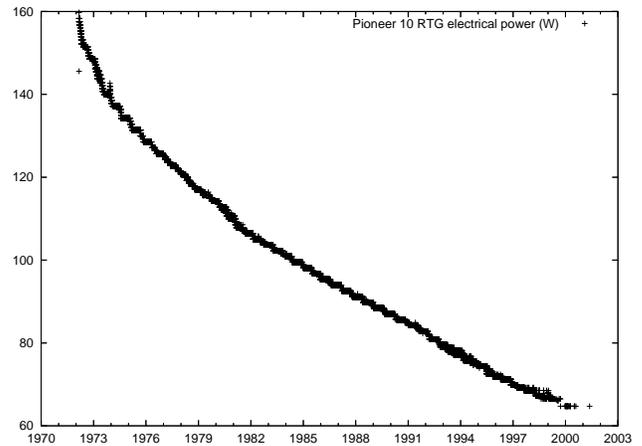, width=\linewidth}
\caption{Changes in total RTG electrical output (in~W) on board Pioneer, as computed using the mission's on-board telemetry.}
\label{fig:elec}
\end{figure}

\begin{figure*}
\includegraphics[width=0.95\linewidth]{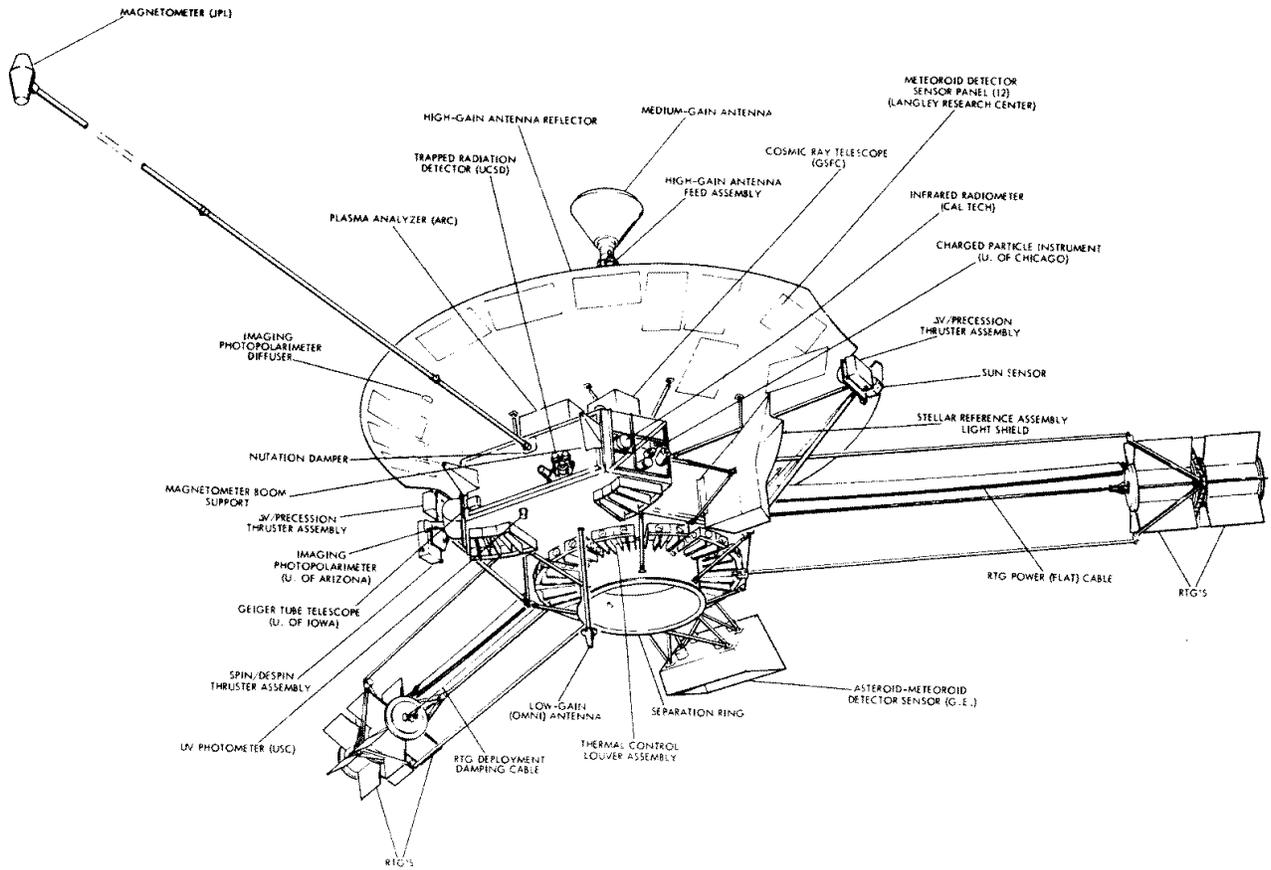}
\caption{A drawing of the Pioneer spacecraft.}
\label{fig:pioneer}
\end{figure*}

The anomalous acceleration of the Pioneer spacecraft is very small. Only $\sim$65~W of anisotropically emitted thermal power is sufficient to produce an acceleration of comparable magnitude. As the Pioneer spacecraft utilized radioisotope thermoelectric generators (RTGs) that produced several kW of waste heat, an anisotropy as small as $\sim$3\% in the pattern of thermal radiation can produce acceleration of the needed magnitude. For this reason, it is very important to characterize the thermal emissions of the Pioneer spacecraft as accurately as possible.

A unique feature of the current effort is the use of telemetry files documenting the thermal and electrical state of the spacecraft. This information was not available previously; however, by May 2006, the telemetry files for the entire durations of both missions were recovered, pre-processed and are ready for the upcoming study. Both of the newly assembled data sets are pivotal to establishing the origin of the detected signal.

All transmissions of both Pioneer spacecraft, including all engineering telemetry, were archived \citep{MDR2005} in the form of files containing Master Data Records (MDRs). Originally, MDRs were scheduled for limited retention. Fortunately, the Pioneers' mission records avoided this fate: with the exception of a few gaps in the data \citep{MDR2005} the entire mission record has been saved, comprising $\sim$60,000 data points for Pioneer 10, and $\sim$50,000 data points for Pioneer 11, a total of $\sim$35~GB. These recently recovered telemetry readings are important in reconstructing a complete history of the thermal, electrical, and propulsion systems for both spacecraft. This, it is hoped, may in turn lead to a better determination of the spacecrafts' acceleration due to on-board systematic effects.

Telemetry formats can be broadly categorized as science formats versus engineering formats. Telemetry words included both analog and digital values. Digital values were used to represent sensor states, switch states, counters, timers, and logic states. Analog readings, from sensors measuring temperatures, voltages, currents and more, were encoded using 6-bit words. This necessarily limited the sensor resolution and introduced a significant amount of quantization noise. Furthermore, the analog-to-digital conversion was not necessarily linear; prior to launch, analog sensors were calibrated using a fifth-order polynomial. Calibration ranges were also established; outside these ranges, the calibration polynomials are known to yield nonsensical results.

With the help of the information contained in these words, it is possible to reconstruct the history of RTG temperatures and power, radio beam power, electrically generated heat inside the spacecraft, spacecraft temperatures, and propulsion system.

\subsection{RTG temperatures and power}

The exterior temperatures of the RTGs were measured by one sensor on each of the four RTGs: the so-called ``fin root temperature'' sensor. Figure~\ref{fig:C201} depicts the evolution of the RTG 1 fin root temperature for Pioneer 10.

A best fit analysis confirms that the RTG temperature indeed evolves in a manner consistent with the radioactive decay of the nuclear fuel on board. The results for all the other RTGs on both spacecraft are similar, confirming that the RTGs were performing thermally in accordance with design expectations.

RTG electrical power can be estimated using two sensor readings per RTG, measuring RTG current and voltage, from which power can be computed by direct calculation.

All this electrical power is eventually converted to waste heat by the spacecrafts' instruments, with the exception of power radiated away by transmitters.

\subsection{Electrically generated heat}

Whatever remains of electrical energy (Fig.~\ref{fig:elec}) after accounting for the power of the transmitted radio beam is converted to heat on-board. Some of it is converted to heat outside the spacecraft body.

The Pioneer electrical system is designed to maximize the lifetime of the RTG thermocouples by ensuring that the current draw from the RTGs is always optimal. This means that power supplied by the RTGs may be more than that required for spacecraft operations. Excess electrical energy is absorbed by a shunt circuit that includes an externally mounted radiator plate. Especially early in the mission, when plenty of RTG power was still available, this radiator plate was the most significant component external to the spacecraft body that radiated heat. A specific telemetry word tells us the shunt circuit current, from which the amount of power dissipated by the external radiator can be computed using the known ohmic resistance of the radiator plate.

Other externally mounted components that consume electrical power are the Plasma Analyzer, the Cosmic Ray Telescope, and the Asteroid/Meteoroid Detector. Though these instruments' exact power consumption is not telemetered, we know their average power consumption from design documentation, and the telemetry bits tell us when these instruments were powered.

Two additional external loads are the battery heater and the propellant line heaters. The power state of these loads is not telemetered. According to mission logs, the battery heater was commanded off on both spacecraft on 12 May 1993 \cite{PNAGENDA93}.

Yet a further external load is the set of cables connecting the RTGs to the inverters. The resistance of these cables is known. Using the RTG current readings it is possible to accurately determine the amount of power dissipated by these cables in the form of heat.

After accounting for all these external loads, whatever remains of the available electrical power on board is converted to heat inside the spacecraft. So long as the body of the spacecraft is in equilibrium with its surroundings, heat dissipated through its walls has to be equal to the heat generated inside.

\subsection{Compartment temperatures and thermal radiation}

\begin{figure}
\includegraphics[width=\linewidth]{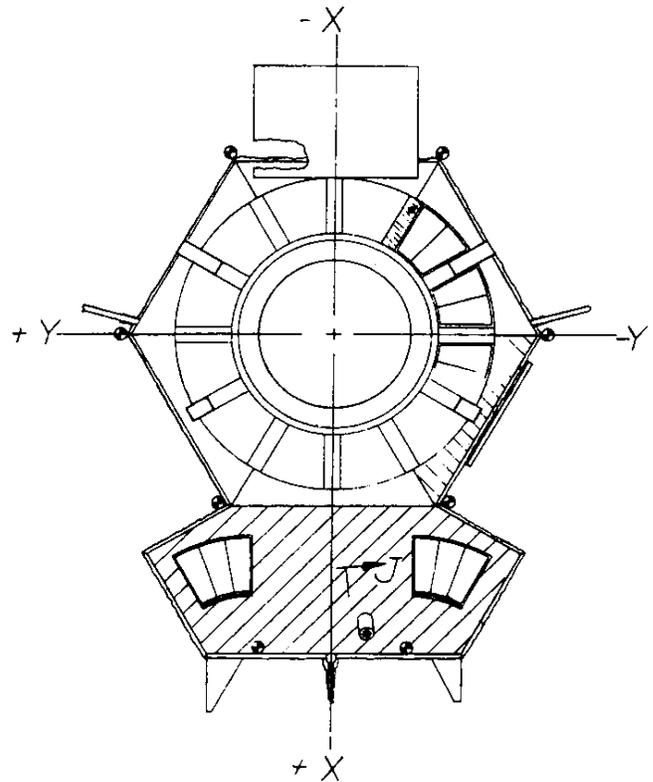}
\caption{Bottom view of the Pioneer 10/11 vehicle, showing the louver system. A set of 12 2-blade louver assemblies cover the main compartment in a circular pattern; an additional two 3-blade assemblies cover the compartment with science instruments.}
\label{fig:louvers}
\end{figure}

\begin{figure}
\includegraphics[width=0.85\linewidth]{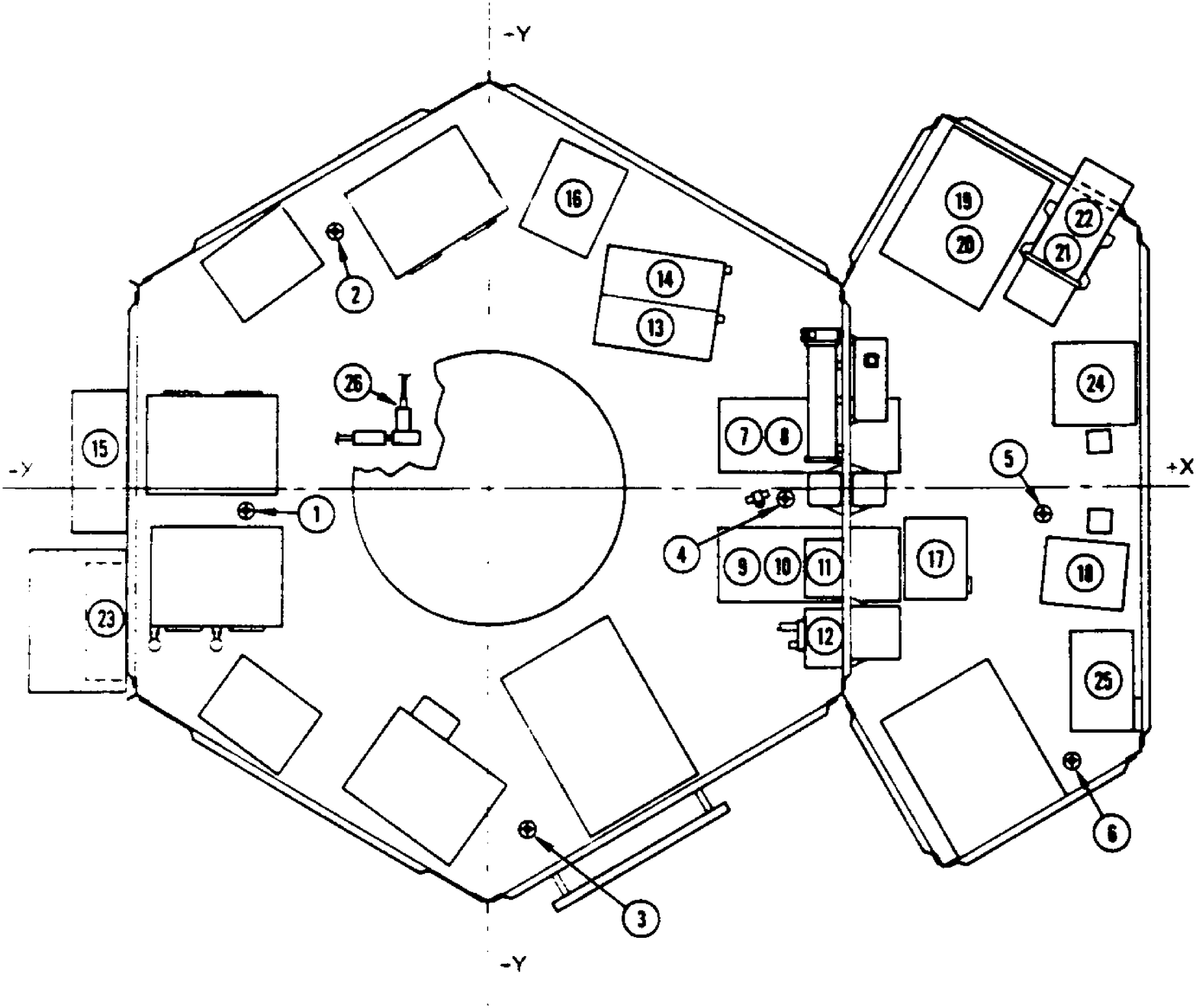}
\caption{Location of thermal sensors in the instrument compartment of Pioneer 10/11 \citep{PC202}. Temperature sensors are mounted at locations 1 to 6.}
\label{fig:tempsens}
\end{figure}

As evident from Fig.~\ref{fig:pioneer}, the appearance of the Pioneer spacecraft is dominated by the 2.74~m diameter high gain antenna (HGA). The spacecraft body, located behind the HGA, consists of a larger, regular hexagonal compartment housing the propellant tank and spacecraft electronics; an adjacent, smaller compartment housed science instruments. The spacecraft body is covered by multilayer thermal insulating blankets, except for a louver system located on the side opposite the HGA, which was activated by bimetallic springs to expel excess heat from the spacecraft.

Each spacecraft was powered by four radioisotope thermoelectric generators (RTGs) mounted in pairs at the end of two booms, approximately three meters in length, extended from two sides of the spacecraft body at an angle of 120$^\circ$. A third boom, approximately 6 m long, held a magnetometer.

The total (design) mass of the spacecraft was $\sim$250~kg at launch, of which 27~kg was propellant \citep{PC202}.

For the purposes of attitude control, the spacecraft were designed to spin at the nominal rate of 4.8~rpm. Six small monopropellant (hydrazine) thrusters, mounted in three thruster cluster assemblies, were used for spin correction, attitude control, and trajectory correction maneuvers (see Fig.~\ref{fig:pioneer}).

The passive thermal control system consisted of a series of spring-activated louvers (see Fig.~\ref{fig:louvers}). The springs were bimetallic, and thermally (radiatively) coupled to the electronics platform beneath the louvers. The louver blades were highly reflective in the infrared. The assembly was designed so that the louvers fully open when temperatures reach 30$^\circ$C, and fully close when temperatures drop below 5$^\circ$C.

As the total exterior area of the spacecraft body and the effective emissivity of the thermal blankets covering the spacecraft are known \cite{SJ1972}, a crude initial calculation can be made determining the amount of heat radiated by the spacecraft walls and through the louver system. These estimates turn out to be consistent with pre-launch thermal vacuum chamber tests \cite{TCSDR3,PFG100151} of the louver system, and with temperature readings obtained from on-board temperature sensors.

There are 6 platform temperature sensors (Fig.~\ref{fig:tempsens}) inside the spacecraft body: 4 are located inside the main compartment, 2 sensors are in the science instrument compartment. The main compartment has a total of 12 2-blade louver blade assemblies; the science compartment has 2 3-blade assemblies.

The thermal vacuum chamber tests provide values for emitted thermal power per louver assembly as a function of the temperature of the electronics platform behind the louver. This allows us to estimate the amount of thermal power leaving the spacecraft body through the louvers, as a function of platform temperatures \citep{MDR2006}, providing means to estimate the amount of heat radiated by the louver system.

\subsection{\label{sec:spin}Spin}

\begin{figure*}[ht]
\hskip -6pt
\begin{minipage}[b]{.5\linewidth}
\centering \psfig{file=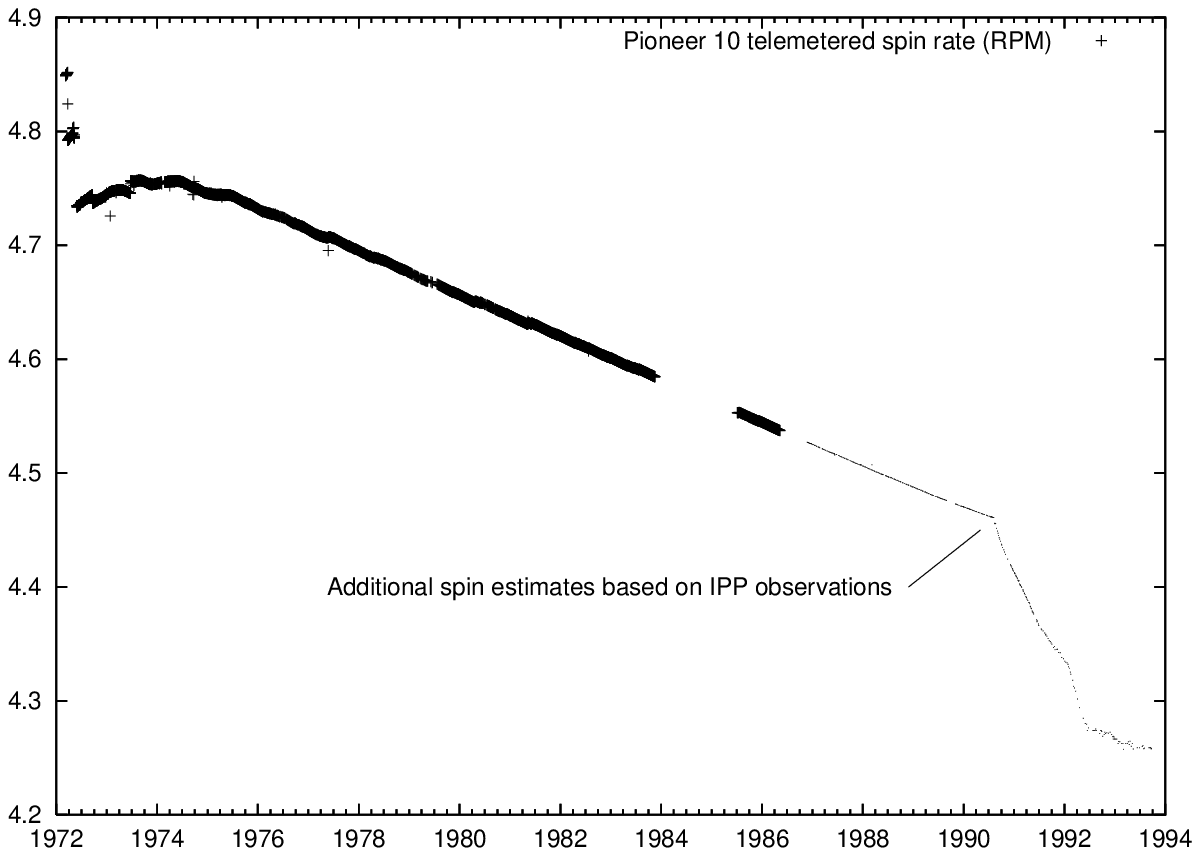, width=\linewidth}
\end{minipage}
\hskip 0.001\linewidth
\begin{minipage}[b]{.5\linewidth}
\centering \psfig{file=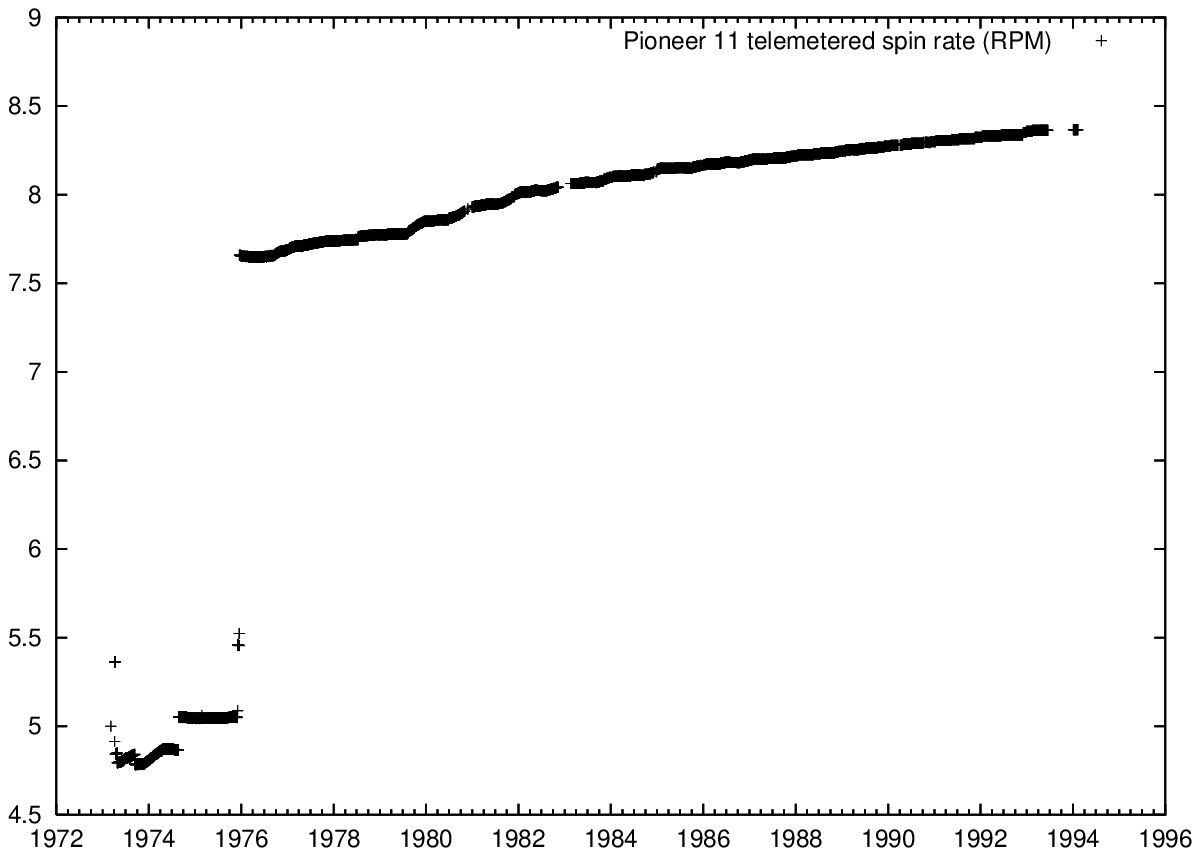, width=\linewidth}
\end{minipage}
\caption{On-board spin rate measurements (in RPM) for Pioneer 10 (left) and Pioneer 11 (right). The sun sensor used on Pioneer 10 for spin determination was temporarily disabled between November 1983 and July 1985, and was turned off in May 1986. Continuing spot measurements of the spin rate were made using the Imaging Photo-Polarimeter (IPP) until 1993. The anomalous increase in Pioneer 11's spin rate early in the mission was due to a failed spin thruster.
}
\label{fig:spin}
\end{figure*}

\begin{figure*}
\hskip -6pt
\begin{minipage}[b]{.5\linewidth}
\centering \psfig{file=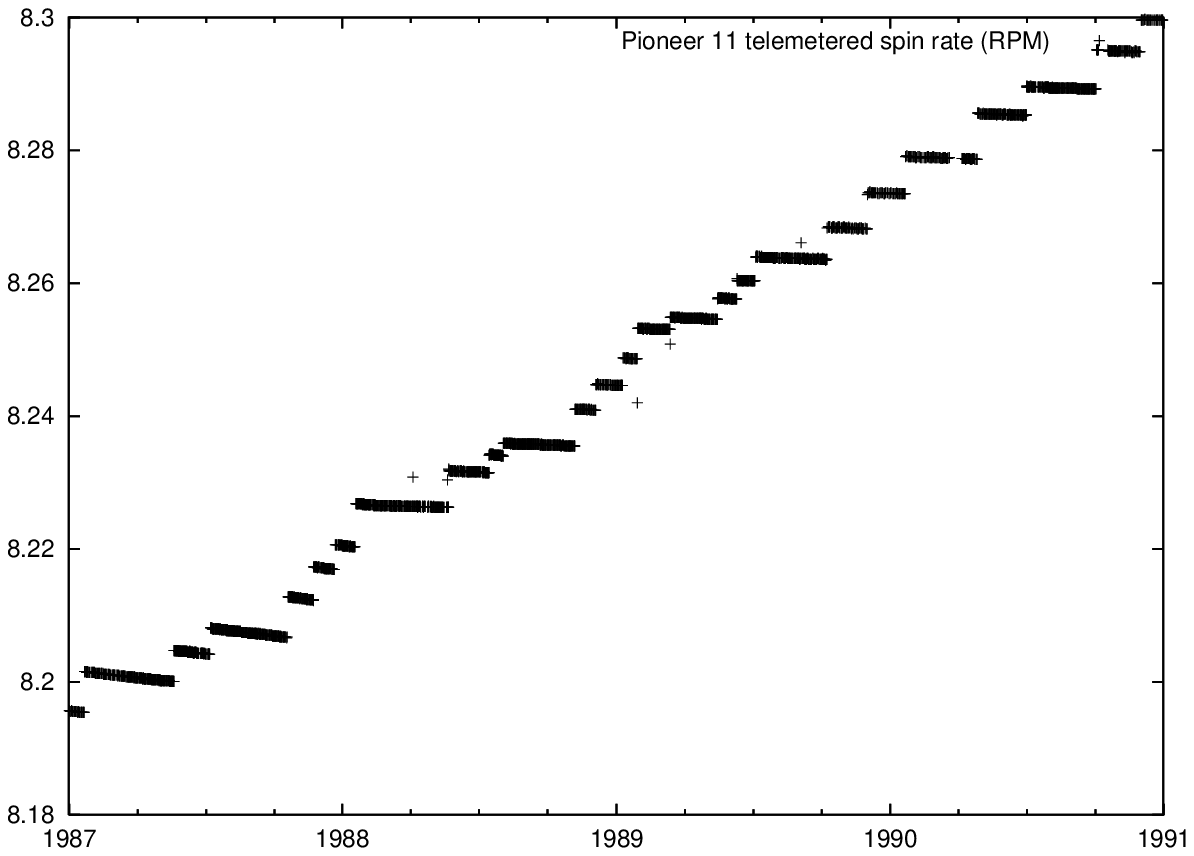, width=\linewidth}
\end{minipage}
\hskip 0.001\linewidth
\begin{minipage}[b]{.5\linewidth}
\centering \psfig{file=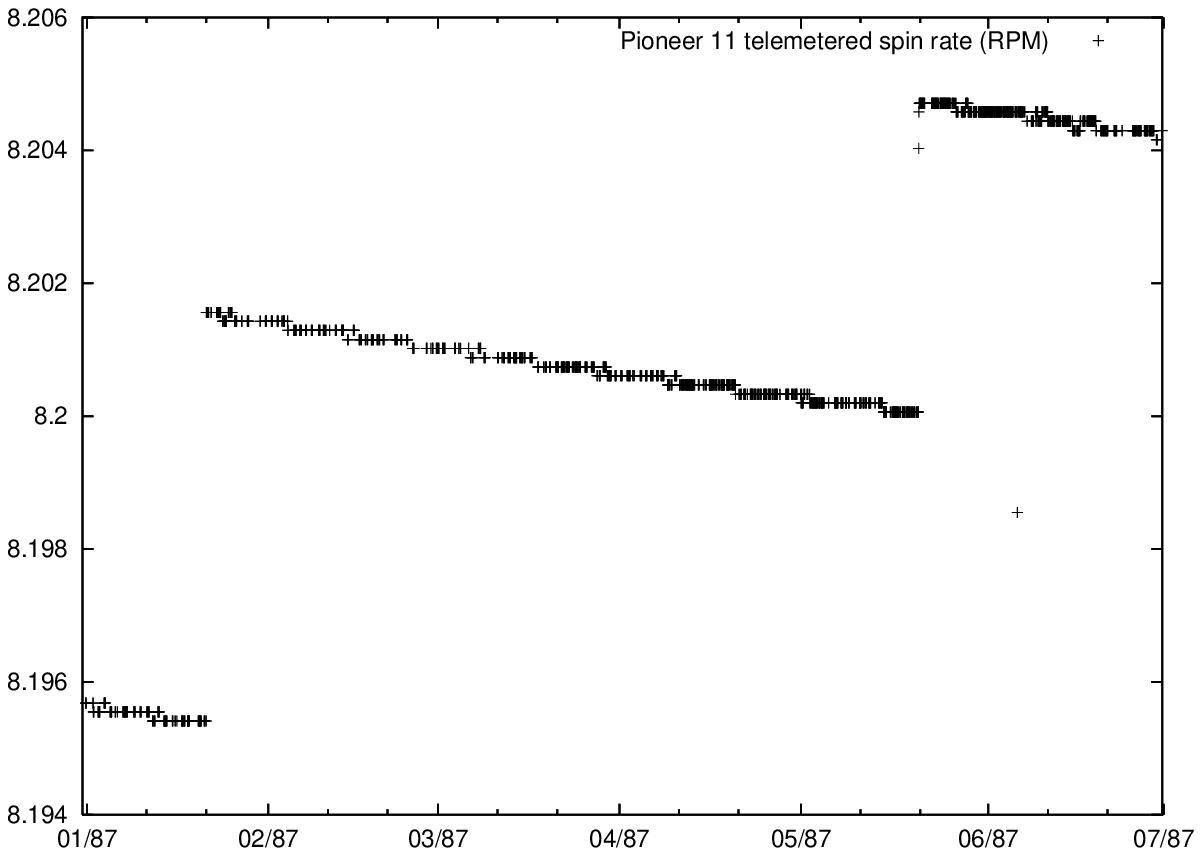, width=\linewidth}
\end{minipage}
\caption{Zoomed plots of the spin rate of Pioneer 11. On the left, the interval examined in ref.~\cite{JPL2002} is shown; maneuvers are clearly visible, resulting in discrete jumps in the spin rate. The figure on the right focuses on the first half of 1987; the decrease in the spin rate when the spacecraft was undisturbed is clearly evident.}
\label{fig:spinzoom}
\end{figure*}

The spin rate of the Pioneer spacecraft was measured independently by three instruments: a star sensor and two Sun sensors. Each of these instruments was capable of providing a roll reference pulse, which in turn was used to synchronize other spacecraft activities.

The spin rate was telemetered to the ground using 3 6-bit words. The resulting 18-bit value measured the time of one full revolution of the spacecraft in units of 1/8192 seconds using an on-board crystal oscillator as a reference clock.

Unfortunately, the star sensor on board Pioneer 10 stopped functioning shortly after Jupiter encounter. The Sun sensors, in turn, could only operate if the Sun was bright enough and appeared sufficiently far from the spacecraft spin axis. Consequently, the Sun sensors stopped providing a useful roll reference pulse after the spacecraft reached a heliocentric distance of $\sim$35 astronomical units.

The Pioneer support team devised an alternate procedure, utilizing the Imaging Photo-Polarimeter (IPP) instrument and its ability to image faint objects in a precisely timed manner \cite{AIAA870502}. Later, when there was no longer sufficient power available on board to operate this instrument, the spin rate was estimated using other means, such as relying on the previously established rate of spin change, and on the spin introducing a slight modulation of the spacecraft's radio signal when the spin axis did not exactly coincide with the spacecraft-Earth line.

The star sensor and Sun sensors remained fully operational on Pioneer 11 throughout its mission.

As a result, we have high accuracy spin information for both spacecraft for much of their missions. The spin behavior of the two spacecraft was not identical (Fig.~\ref{fig:spin}). The spin rate of Pioneer 10 was decreasing steadily, and does not appear to be affected by maneuvers. The spin rate of Pioneer 11 increased rapidly early in the mission as a result of a spin thruster malfunction, but it continued to increase afterwards. A close examination of the spin behavior reveals, however, that the increases in spin rate are discontinuous and coincide with maneuvers; between maneuvers, the spin rate was decreasing (Fig.~\ref{fig:spinzoom}).

\subsection{Maneuvers}

Each of the Pioneer spacecraft was equipped with six monopropellant hydrazine thrusters, designed to perform three types of maneuvers: spin-up/spin-down maneuvers, delta-$V$ maneuvers, and precession maneuvers. The first of these was used early in their missions to adjust the spin rate after the deployment of the RTG and magnetometer booms. The second type of maneuver was used to correct the spacecrafts' trajectories, to achieve the desired planetary encounters with Jupiter and (for Pioneer 11) Saturn. Precession maneuvers, in turn, were used regularly throughout the mission; their goal was to adjust the spacecrafts' spin axis, to ensure that the HGA continuously points in the direction of the Earth.

In theory, precession maneuvers do not alter the velocity of the spacecraft, only change their orientation. In practice, due to small differences between thrusters, uncertainties in the duration of firing pulses, thruster misalignment, and outgassing after maneuvers, a small, essentially random change in the spacecrafts' motion is expected. These changes can be modeled, for instance, as an instantaneous change in the line-of-sight velocity of the spacecraft. However, before one can do such modeling, the time when the maneuvers occurred must be known.

\begin{figure*}[ht!]
\includegraphics[width=1.0\linewidth]{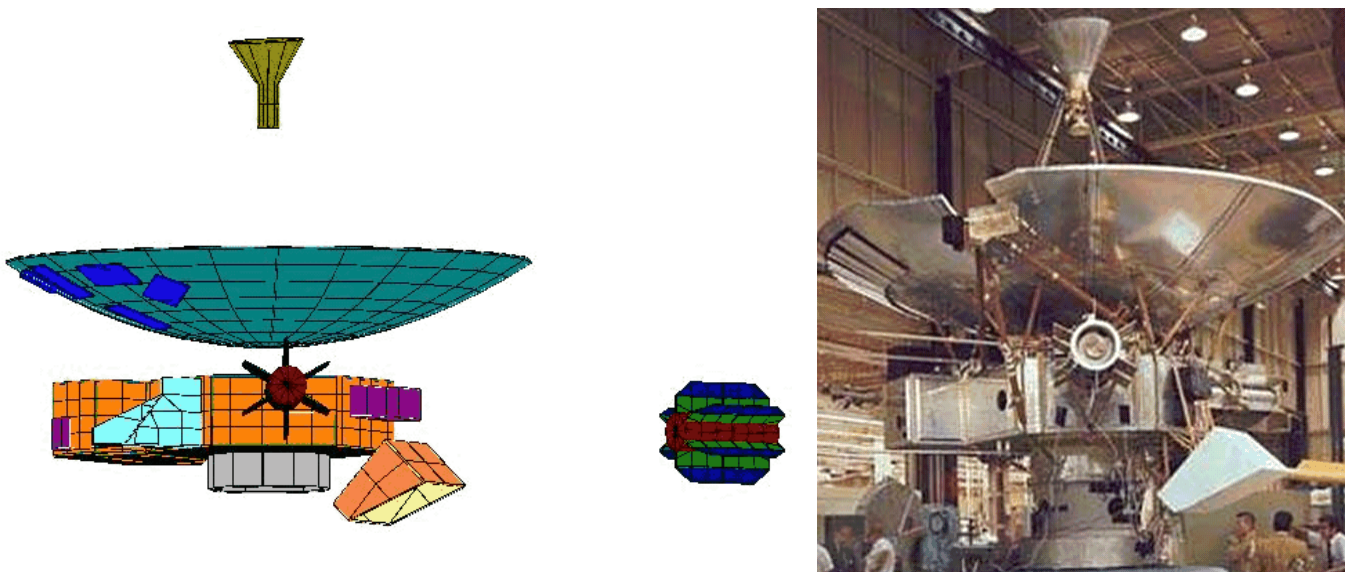}
\caption{A geometric model (left) of the Pioneer spacecraft, used for finite element analysis, and a photograph (right) of Pioneer 10 prior to launch. The geometric model accurately incorporates details such as the Medium Gain Antenna (MGA), the Asteroid-Meteoroid Detector, and the star sensor shade. Note that in the geometric model, the RTGs are shown in the extended position; in the photograph, the RTGs are stowed.}
\label{fig:geometry}
\end{figure*}

Flight telemetry contains information about the number of pulses fired by each of the six thrusters on board. The number of firing pulses recorded, and the thrusters participating in the maneuver unambiguously identify the type of maneuver that was performed.

Although thruster pulse counts are monitored continuously, due to low data rates, the corresponding data words are telemetered as infrequently as once every 51.2 minutes at the lowest telemetry data rate. Due to gaps in ground station coverage or missing data records, there are instances when an increase in the thruster pulse count is bracketed by two telemetry readings that are many hours, perhaps even more than a day apart. However, due to the smallness of the velocity change as a result of a precession maneuver, even such an uncertainty in the timing of the maneuver does not adversely affect the accuracy of orbital modeling.

\section{\label{sec:rec-force}Modeling the thermal recoil force}

Even with all the recently recovered telemetry information, modeling the thermal recoil force on the Pioneer spacecraft remains a difficult task. The main reason for this can be summed up as follows: What we seek is not just the amount of heat emitted by the spacecraft (which we know with good precision), but the small anisotropy in the pattern of its thermal emissions, an inherently second-order effect.

The traditional way of estimating thermal recoil forces is to build a detailed thermal model using, for instance, finite element analysis tools. We are also exploring an alternative approach that combines the use of flight telemetry with precision orbit calculations, and provides a more direct and, it is hoped, more accurate estimate of the anisotropy of thermal radiation \cite{t-model-2007}.

\subsection{Finite element thermal modeling}

\begin{figure*}[t!]
\includegraphics[width=0.85\linewidth]{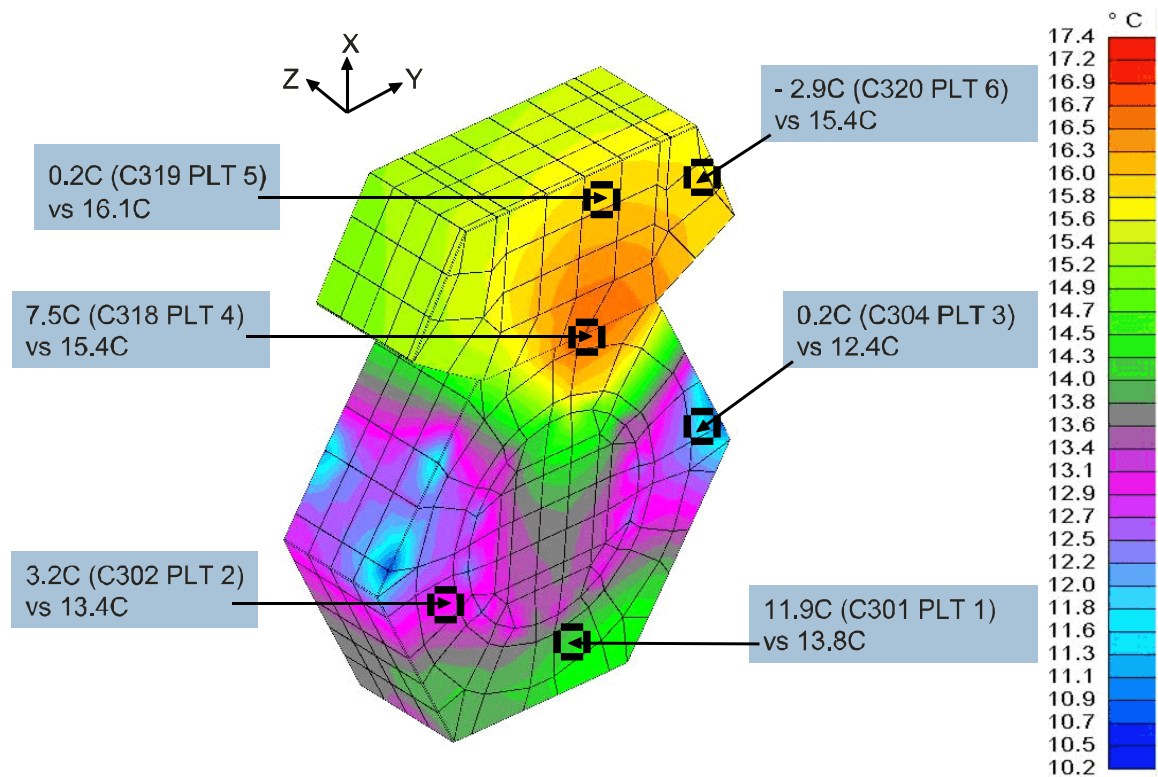}
\caption{A ``work-in-progress'' temperature map of the outer surface of the Pioneer 10 spacecraft body, comparing temperatures calculated via a numerical finite element method vs. temperatures measured by platform temperature (PLT) sensors and telemetered. While agreement between calculated and telemetered temperatures is expected to improve as the model is being developed, discrepancies between these values illustrate the difficulties of creating a reliable temperature map using numerical methods.}
\label{fig:tempmap}
\end{figure*}

The radiative exchange of thermal energy between two surfaces is described by the equation
\begin{equation}
Q=\int_{A_2}\int_{A_1}\frac{I\cos{\theta_1}\cos{\theta_2}}{r^2}~dA_1dA_2.
\label{eq:rad}
\end{equation}
where $Q$ is the radiative power emitted by the surface $A_1$ that is absorbed by the surface $A_2$, $I$ is the intensity of radiation at surface element $dA_1$, $r$ is the distance between surface elements $dA_1$ and $dA_2$, and $\theta_1$ and $\theta_2$ are the angles between the line connecting surface elements $dA_1$ and $dA_2$ and their respective normals.

The intensity of radiation is related to the radiant emittance $q$ by the formula $I=q/\pi$. The radiant emittance, in turn, can be calculated from the surface temperature $T$ and emissivity coefficient $\epsilon$ using the Stefan-Boltzmann law:
\begin{equation}
q=\sigma\epsilon T^4,
\end{equation}
where $\sigma\simeq 5.67\times 10^{-8}$~Js$^{-1}$m$^{-2}$K$^{-4}$ is the Stefan-Boltzmann constant.

In principle, these equations completely determine the thermal emissions of an object. The anisotropy of thermal emissions can be determined, for instance, by surrounding the object with an infinite (or, at least, sufficiently large) black sphere and calculating the amount of radiation absorbed by this sphere using Eq.~(\ref{eq:rad}).

In order to carry out the calculations, however, one must have accurate knowledge of
\begin{itemize}
\item the emitting object's detailed geometry;
\item the object's temperature at all points across its surface;
\item the infrared emissivity coefficient $\epsilon$ at all points across the surface, and its dependance on the temperature.
\end{itemize}

This information is not always readily available. In the case of Pioneer 10 and 11, only a few detailed drawings have been found from which the spacecraft's geometry could be estimated. Surface properties are not known very accurately, and very little is known about the effects of aging in the deep space environment. Although the spacecraft had several temperature sensors on board, these were designed to measure interior temperatures; the external temperature distribution can only be calculated by constructing a complex model that takes into account the internal structure and heat conductive properties of the spacecraft.

Despite these difficulties, a highly detailed mathematical model of the spacecraft has been constructed successfully \cite{THERMSTAT}. This model is used to calculate radiative heat exchange between spacecraft surfaces, absorbed solar loads, heat flows, and predicted temperatures. The finite element model incorporates $\sim$3000 nodes and 2600 plate elements, using 3.4 million radiation conductors and $\sim$7000 linear conductors.

\begin{figure}
\includegraphics[width=0.9\linewidth]{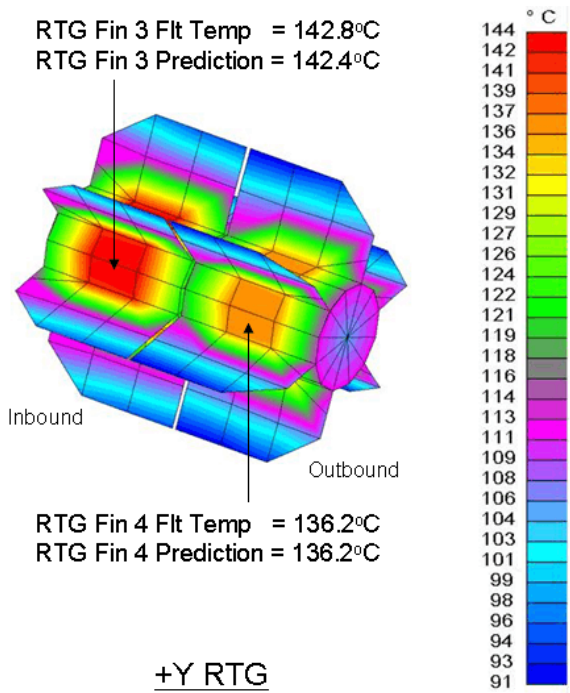}
\caption{Modeling the exterior temperatures of a pair of RTGs using finite element software. Predicted values are within a few tenth of a degree from temperature readings obtained from flight telemetry.}
\label{fig:rtgtemp}
\end{figure}

Results from this model can be validated against the readings from the spacecraft's temperature sensors, as seen in Figs.~\ref{fig:tempmap} and \ref{fig:rtgtemp}.

Flight telemetry can provide another important means of model verification. In addition to temperature sensors, flight telemetry also offers readings from which the thermal power generated on board can be calculated. The principle of energy conservation dictates that in a spacecraft that's in steady state, the amount of thermal power generated must equal the amount of heat emitted by the spacecraft. Because of this, temperature and power readings can be viewed as redundant parameters characterizing the same physical processes.

\subsection{Estimating thermal recoil forces from orbital analysis}

Our analysis of the geometry and thermal properties of the Pioneer spacecraft yielded the following findings:
\begin{itemize}
\item The thermal output of the Pioneer spacecraft can be accurately modeled using as few as two heat sources, electrical and RTG heat;
\item As the spacecraft is spinning, lateral forces result in no significant long-term acceleration; the spin axis, in turn, points approximately in the direction of the Earth at all times;
\item The acceleration of the spacecraft due to thermal radiation is a linear function of the power of the internal power sources.
\end{itemize}

In other words, the thermal acceleration of the Pioneer 10 and 11 spacecraft due to heat anisotropically rejected off the vehicles can be modeled by a simple equation (see details in \cite{t-model-2007}):
\begin{equation}
\vec{a}_\mathrm{thermal}=\frac{2}{3mc}\sum_i\xi_iP_i,
\label{eq:thermal}
\end{equation}
where $c$ is the speed of light, $m$ is the spacecraft's mass, $P_i$ is the $i$-th heat source, and $\xi_i$ is the associated dimensionless efficiency factor. (The additional factor of $2/3$ arises as a result of modeling the spacecraft's exterior surfaces as Lambertian emitters.)

\begin{figure}
\includegraphics[width=\linewidth]{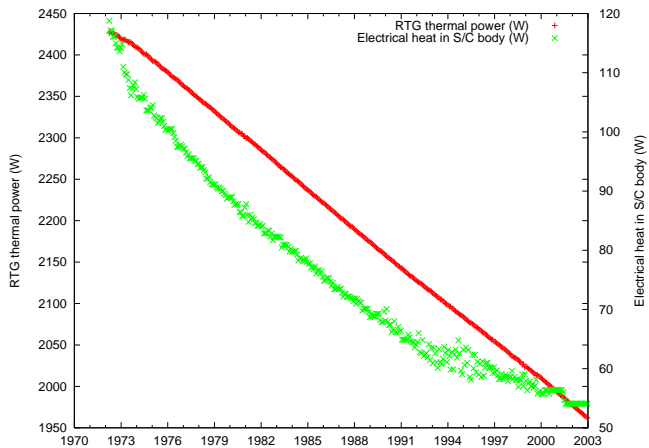}
\caption{Heat generated by RTGs (red, approximately straight line) and electrical equipment (green) in Pioneer 10 over the lifetime of the spacecraft.}
\label{fig:power}
\end{figure}

The values of $P_i$ are known from flight telemetry or design documentation (Fig.~\ref{fig:power}). The values of $\xi_i$ can be determined approximately by analyzing the spacecraft's geometry and surface properties, as outlined in the previous section.

The simplicity of the relationship between heat and acceleration suggests that another approach may be possible. Eq.~(\ref{eq:thermal}) may be incorporated directly into the equations of motion, allowing one to designate the efficiency factors $\xi_i$ as parameters to be fitted. This, radically different approach requires no knowledge of the geometry or thermal properties of the spacecraft. Question is, does it yield believable results?

Recently, one of us (VTT) constructed an orbit determination program specifically to incorporate into the equations of motion Eq.~(\ref{eq:thermal}), with the values of $P_i$ supplied directly from flight telemetry. The software allows us to test a variety of hypotheses, attributing some, or all, of the anomalous acceleration to thermal radiation, but also incorporating other fictitious forces. This program could also be used to guide our investigation with JPL's Orbit Determination Program, similar to the earlier efforts in the study of the effect (see discussion in \cite{JPL1998,JPL2002,JPL2005}).

This work is aimed to answer three questions:
\begin{itemize}
\item Does such an approach yield values for the efficiency factors $\xi_i$ that are consistent with the values calculated using conventional methods?
\item Is it possible to obtain a good orbital solution incorporating the thermal recoil force, without resorting to the use of other forces?
\item Is it possible to distinguish between solutions with a thermal recoil force vs. solutions that incorporate other forces, such as a constant acceleration pointing towards the Sun?
\end{itemize}

The results of this work to date, while not conclusive, are encouraging. Preliminary estimates of $\xi_i$ are consistent with the results obtained from finite-element thermal analyses, including the sophisticated on-going efforts at JPL and our earlier, simpler numerical integrations. It is our hope that the two approaches will complement each other, and that the accuracy necessary to determine the extent to which anisotropic thermal radiation from the spacecraft can be responsible for the anomalous acceleration (i.e., an acceleration accuracy of $\sim 10^{-10}$~m/s$^2$ will be achievable.

\section{\label{sec:summ}Conclusions}

By 2007, the existence of the Pioneer anomaly is no longer in doubt. A steadily growing part of the community has concluded that the anomaly should be subject to further investigation and interpretation. Our continuing effort to process and analyze Pioneer radio-metric and telemetry data is part of a broader strategy (see discussion in \citep{JPL2005,MDR2005}).

Based on the information obtained from flight telemetry, we were able to develop a high accuracy thermal, electrical, and dynamical model of the Pioneer spacecraft. This model will be used to further improve our understanding of the anomalous acceleration and especially to study the contribution from the on-board thermal environment to the anomaly.

It is clear that a thermal model for the Pioneer spacecraft would have to account for all heat radiation produced by the spacecraft. One can use telemetry information to accurately estimate the amount of heat produced by the spacecrafts' major components. The next step is to utilize this result along with information on the spacecrafts' design to estimate the amount of heat radiated in various directions.

This entails, on the one hand, an analysis of all available radio-metric data, to characterize the anomalous acceleration beyond the periods that were examined in previous studies. Telemetry, on the other hand, enables us to reconstruct a thermal, electrical, and propulsion system profile of the spacecraft. Soon, we should be able to estimate effects due to on-board systematic acceleration sources, expressed as a function of telemetry readings. This provides a new and unique way to refine orbital predictions and may also lead to an unambiguous determination of the origin of the Pioneer anomaly.

\begin{acknowledgments}

We would like to express our gratitude to our many colleagues who have either collaborated with us on this manuscript or given us their wisdom. We specifically thank Sami Asmar, Gene L. Goltz, Kyong J. Lee, Timothy P. McElrath, and Neil Mottinger of JPL for their help in understanding and conditiononing of the Pioneer Doppler data. Our gratitude also goes to Gary Kinsella of JPL and Louis K. Scheffer of Cadence Design Systems who benefited us with their insightful comments and suggestions regarding thermal modeling of the Pioneers.

Part of this work was initiated during our visit to the Perimeter Institute for Theoretical Physics, Waterloo, Canada. In this respect we would like to thank John Moffat for his hospitality and support. We also thank The Planetary Society for support and, in particular, Louis D. Freidman, Charlene M. Anderson, and Bruce Betts for their interest, stimulating conversations and encouragement. The work described here, in part, was carried out at the Jet Propulsion Laboratory, California Institute of Technology, under a contract with the National Aeronautics and Space Administration.

~\par
~\par

\end{acknowledgments}

\bibliographystyle{aipproc}   

\end{document}